# Optical holonomic single quantum gates with a geometric spin under a zero field

Yuhei Sekiguchi, Naeko Niikura, Ryota Kuroiwa, Hiroki Kano and Hideo Kosaka[*]

Yokohama National University, 79-5 Tokiwadai, Hodogaya, Yokohama 240-8501, Japan

*kosaka-hideo-yp@ynu.ac.jp

# Abstract

Realization of fast fault-tolerant quantum gates on a single spin is the core requirement for solid-state quantum-information processing. As polarized light shows geometric interference, spin coherence is also geometrically controlled with light via the spin-orbit interaction. Here, we show that a geometric spin in a degenerate subspace of a spin-1 electronic system under a zero field in a nitrogen vacancy center in diamond allows implementation of optical non-adiabatic holonomic quantum gates. The geometric spin under quasi-resonant light exposure undergoes a cyclic evolution in the spin-orbit space, and acquires a geometric phase or holonomy that results in rotations about an arbitrary axis by any angle defined by the light polarization and detuning. This enables universal holonomic quantum gates with a single operation. We demonstrate a complete set of Pauli quantum gates using the geometric spin preparation and readout techniques. The new scheme opens a path to holonomic quantum computers and repeaters.

# Main text

A quantum bit or qubit must be capable of being precisely and quickly manipulated, as well as robust against noise. These criteria pose a dilemma in that the qubit must be open for a driving field but not for a noise field. It has been demonstrated that the degenerate subspace of a spin-1 electronic system under a zero field, which we call a geometric spin, can serve as a promising memory qubit that is robust against environmental noise[1]. The challenge is to manipulate the degenerate qubit with the help of a geometric phase.

The concept of the geometric phase was first proposed by Pancharatnam in 1956[2] in reference to light polarization. Since then, two kinds of geometric phase have been discussed. Adiabatic geometric phases were first proposed by Berry in 1984[3], and non-adiabatic non-Abelian geometric phases were proposed by Anandan in 1988[4]. Holonomic quantum computation (HQC) based on the adiabatic geometric phase was then proposed for fault-tolerant quantum gates by Zanardi and Rasetti in 1999[5], and generalized to non-adiabatic HQC by Wang and Matsumoto in 2001[6,7] and Zhu and Wang in 2002[8]. The geometric phase has been experimentally demonstrated in molecular ensembles[8,9], in a superconducting qubit[10], in trapped ions[11,12], in a quantum dot[13,14], and in a single nitrogen-vacancy (NV) center in diamond[15–17].

These demonstrations, however, introduced fundamentally unnecessary energy splitting to the qubit to energetically select a well-defined eigenstate, or to utilize spin precession for arbitrary axis rotation. The splitting in turn prevented fast rotation, even in the non-adiabatic case, to avoid unwanted interference between the eigenstates. To overcome this problem, Sjöqvist[18,19] proposed an optical scheme with a three-level Λ system to implement non-adiabatic HQC based on Anandan's scheme[4], although the importance of degeneracy in the qubit space was not mentioned. Meanwhile, it was experimentally demonstrated that degeneracy plays a key role in the preparation[20] and readout[21] of a geometric spin in any arbitrary state defined by light polarization, both to detect quantum entanglement between a photon and geometric spin[22], and to transfer the quantum state of a photon to a geometric spin[23]. More importantly, it was also shown that degeneracy of the spin basis states under a zero field allows the time-inversion symmetry of the system to maintain spin coherence[1]. Here, we demonstrate the holonomic quantum gate of a geometric spin with geometric rotation about an arbitrary axis by any angle defined by the light polarization, and detuning using the degenerate three-level Λ system in a negatively charged NV center in diamond.

An NV center in diamond (Fig.1a) offers an ideally degenerate three-level Λ system under a zero field, where the degenerate $m_S = \pm 1$ states at the orbital ground state $(|0\rangle_L|\pm1\rangle_S$, hereafter denoted as $|\pm1\rangle$), serve as the geometric spin bases (Fig.1b). On the other hand, one of the orbital excited states, called $|A_2\rangle = (|+1\rangle_L|-1\rangle_S + |-1\rangle_L|+1\rangle_S)/\sqrt{2}$ [24], which is known to generate[25] or measure[22] the entanglement of the photon polarization and the electron spin, offers the orbital parameter space or holonomy space of the time-dependent Hamiltonian parameterized by $\theta$ and $\phi$ (Fig.1c). Note that the transition driven by microwave between the basis states of the geometric spin $|\pm1\rangle$, is prohibited since it requires double transition. A classical light for rotation in the polarization state $|\psi\rangle_p = \cos\left(\frac{\theta}{2}\right)|+1\rangle_p + e^{i\phi}\sin\left(\frac{\theta}{2}\right)|-1\rangle_p$ based on circular polarizations $|\pm1\rangle_p$ couples the $|A_2\rangle$ with the corresponding state called the bright state $|B\rangle = \sin\left(\frac{\theta}{2}\right)|+1\rangle + e^{i\phi}\cos\left(\frac{\theta}{2}\right)|-1\rangle$, to induce orbital evolution in the spin-orbit space spanned by the $|A_2\rangle$ and $|B\rangle$ states; in contrast, it uncouples the orthogonal state called the dark state $|D\rangle = \cos\left(\frac{\theta}{2}\right)|+1\rangle - e^{i\phi}\sin\left(\frac{\theta}{2}\right)|-1\rangle$, which remains unchanged[22], as described by the parameter-dependent interaction Hamiltonian $\widehat{H} = \frac{\Omega}{2}(|A_2\rangle\langle B|+|B\rangle\langle A_2|) - \Delta|A_2\rangle\langle A_2|$, where $\Omega$ denotes the Rabi frequency on resonance, and $\Delta$ denotes the detuning (Methods). The unitary operator after a cyclic evolution in the spin-orbit space showing the holonomy matrix[19,26] is represented as $U_\Lambda(t_{2\pi}) = e^{-i\gamma}|B\rangle\langle B| + |D\rangle\langle D| = e^{-i\gamma n_B \cdot \sigma}$, where the bright state acquires a geometric phase or holonomy of $\gamma = \pi\left(1 - \frac{\Delta}{\sqrt{\Omega^2+\Delta^2}}\right)$, which corresponds to one-half of the solid angle enclosed by the orbital trajectory in the spin-orbit space (Fig. 1c). The cycle time (Rabi period) is shortened by the detuning as $\frac{2\pi}{\sqrt{\Omega^2+\Delta^2}}$ (Methods). Note that the proper parameter set of $\{\theta, \phi, \Delta\}$ allows the geometric spin to be rotated by any angle $\gamma$ about an arbitrary axis defined by the unit gyration vector $\boldsymbol{n}_B$ pointing to $|B\rangle$, as shown in Fig. 1c. Also note that the unitary operators with different sets of $\{\theta, \phi, \Delta\}$ do not generally commute with each other to satisfy the non-Abelian condition required for implementing universal holonomic gates.

The experimental setup, pulse sequence and photoluminescence spectrum are shown in Figs. 2a, b and c, respectively (see Methods). We begin with preparatory experiments to calibrate the pulse length of the rotation light. The optically-driven Rabi oscillation in the spin-orbit space with vertically polarized light (Fig. 2d) indicates that the geometric spin prepared in the bright state undergoes a cyclic evolution in 1.7 ns, which is well within the damping time of 9 ns, with a rotation light of 30 µW. The Rabi frequency depends on the light polarization due to the off-alignment of the optical axis against the NV axis. The pulse length is thus calibrated depending on light polarization and detuning to provide the expected rotation angle, as shown in Fig. 2e.

We evaluate the quantum process of the optical geometric spin rotation by reconstructing the geometric spin state vectors rotated around the X-, Y- and Z-axes based on quantum state tomography, which are expected to trace the great circles in the Bloch sphere. The reconstructed states, however, largely deviate from the expected trajectories, as shown in Fig. 3a. These state vectors are well-reproduced by the Hamiltonian analysis, including the off-alignment of the NV center and the crystal strain (Methods). The unwanted effects are well-compensated by adapting the light polarization to correct the dark states in order to prepare ideal states and bright states that faithfully rotate and readout the states, as shown in Fig. 3b (Methods).

The fidelities of the optical holonomic quantum gates are evaluated by quantum process tomography (Methods). The process fidelities for the holonomic Pauli-X (bit flip or NOT), Y (bit and phase flip) and Z (phase flip) gates are 92(11)%, 89(3)% and 90(10)%, respectively (Fig. 3c). The fidelities are mainly limited by the dephasing induced by the real excitation to the $A_2$ state followed by the spontaneous relaxation back to the $|\pm1\rangle$ states, seen as amplitude damping of the optical Rabi oscillation in Fig. 2d. Other elemental operations required to build universal quantum gates such as the Hadamard gate $H = \frac{1}{\sqrt{2}}\begin{bmatrix}1 & 1\\1 & -1\end{bmatrix}$, phase gate $S = \begin{bmatrix}1 & 0\\0 & i\end{bmatrix}$, and $\pi/8$-rotation gate $T = \begin{bmatrix}1 & 0\\0 & e^{i\frac{\pi}{4}}\end{bmatrix}$, can also be implemented with a single operation by choosing the proper parameter sets of $\{\theta, \phi, \Delta\} = \{\frac{\pi}{4}, 0, 0\}$ for H, $\{0, 0, \frac{\Omega}{\sqrt{3}}\}$ for S, and $\{0, 0, \frac{3\Omega}{\sqrt{7}}\}$ for T to achieve the shortest pulse length of the rotation light. The fidelities could be further increased by increasing the power to decrease the pulse length up to the limit given by the splitting between the $A_2$ and $A_1$ states (typically around 3 GHz). Although we showed only one-qubit gates, strong dependence on detuning in the nearly resonant scheme, shown in Fig. 2(e), in contrast to the conventional far-detuned or off-resonant scheme[27–29], allows implementation of two-qubit gates, such as the C-Phase gate. For example, if we use a proximal $^{13}$C isotope with hyperfine coupling stronger than the natural linewidth of the $A_2$ state, we can selectively gate the electron spin under the conditions of the nuclear spin, such as $\hat{Z}_S|+1\rangle_I\langle+1| + |-1\rangle_I\langle-1|$, which is simply the C-Phase gate. With those single-qubit and two-qubit gates, we could construct a universal non-Abelian non-adiabatic HQC[5,19].

The arbitrary angle rotation obtained here is extremely important for achieving fast quantum Fourier transform[30], which requires controlled-$R_k$ gates, where $R_k = \begin{bmatrix} 1 & 0 \\ 0 & e^{\frac{2\pi i}{2^N}} \end{bmatrix}$ and $k$ is an integer less than the system size $N$. Although it is possible to construct $R_k$ gates with a combination of {X, H, S, T} gates, this is not efficient for handling large systems. On the other hand, the holonomic quantum gate directly performs the $R_k$ gates.

The geometric rotation on near resonance is essentially different from the conventional spin rotation based on the three-level Λ system with far detuning or off resonance[27–29]. Figure 4a shows simulated gate fidelity for X-gates as a function of detuning and pulse length. When the detuning is smaller than the Rabi frequency of 250MHz, one turn (or a few turns) of cyclic evolution in the spin-orbit space induce the geometric phase in the spin space. In contrast, when the detuning is much larger than the Rabi frequency, the dynamic phase instead of the geometric phase accumulates continuously to result in the spin rotation, which is known as the stimulated Raman transition[27–29] or the optical Stark effect[27–29]. Although the conventional off-resonant scheme seems better for reducing the probability of transition into the excited state and the subsequent spontaneous relaxation, it requires a longer operation time than in the present near-resonant scheme; the latter forces the excited state back to the ground state before relaxation, and thus results in higher fidelity than in the off-resonant stimulated-Raman scheme (Fig. 4a). The operation time is further reduced with a smaller rotation angle, leading to further high fidelity. Note that the gate fidelities for the $2\pi$-rotation shown in Fig. 3c are the worst case in the X-, Y-, and Z-axis. Note also that the geometric rotation is extremely robust against pulse-length error in comparison to the straightforward dynamic rotation in the two-level system, especially for larger angle rotations, as shown in Figs. 4(b) and (c). The detailed tolerance analyses against various types of energy shifts are shown in Supple. Fig. 1. Although some tendencies are seen to depend on rotation angle, the tolerances are generally large enough for unexpected field fluctuation, and more importantly, optical dynamic rotation can never be used for the ground state spin in practice.

The geometric rotation relies only on light polarization, not on the spin precession conventionally used to compose an arbitrary axis rotation. It was performed under a completely zero field, which is desirable for the quantum coherence of the geometric spin, because it means the surrounding $^{13}$C nuclear spins are completely frozen[1]. Moreover, the unwanted timing problem caused by the interference from those spins is avoided. Although detuning was used in the demonstration to change the rotation angle, the same operation could be performed by discreet phase change, instead of the continuous phase change called detuning used here. The typical approach to discreet phase change is to change the phase of the rotation light at the north pole ($|A_2\rangle$) in the spin-orbit space. The general composite pulse scheme ideally relaxes the requirements for specific pulse lengths, and pulse length error would not depend on the rotation angle. However, this scheme lowers the fidelity somewhat, especially for small angle rotation, due to the longer total operation time. Thus we chose the continuous phase change scheme rather than the discreet phase change scheme for faster operation and higher fidelity.

The demonstration by Yale et al.[17] used the dark state in the $|\pm 1\rangle$ space to acquire the geometric phase called the Berry phase[3] by adiabatically changing the dark state to make a cyclic evolution. In contrast, our demonstration used the bright state to acquire the geometric phase called the Aharonov-Anandan phase[4] by non-adiabatically applying a short optical pulse to make a cyclic evolution in the spin-orbit space, thus enabling faster quantum gate operations within a few ns instead of a few hundred ns, as in Yale's demonstration. We also achieved complete rotation about an arbitrary axis by any angle corresponding to the light polarization and the phase, with fidelities as high as 90%, in strong contrast to Yale's experiments.

Combined with our previous demonstrations of optical spin state tomography[22] and optical spin state preparation[2], our present observations established a complete set of elemental quantum operations for a solid-state spin composed of initialization, manipulation, and readout. The developed method could also be applied to the manipulation of a nano-spin to provide an optical pickup for the MRAM. It is also suitable for the ultra-sensitive quantum sensor beyond the classical limit, since it works perfectly under a zero magnetic field.

Geometric rotation was used to implement non-adiabatic holonomic quantum gates to manipulate states of geometric spin qubits, which are known to be robust against control errors and environmental noise,[1,19] and are thus expected to be building blocks of the HQC. Our approach is applicable to any other three-level system based on defect centers, ion traps, quantum dots and superconducting circuits. The optical control is especially useful for individually addressing integrated spins, which require nano-scale local access. It thus opens a path to building holonomic quantum processors for quantum computers. The holonomic quantum gates could also be used for the basis transformation between four Bell states, and optimization or calibration of the teleportation-based quantum state transfer[23] to build the holonomic quantum repeaters used for the long-distance quantum communication network[31].

# Methods

**Experimental Setup.** We used a native NV center in a high-purity type-IIa chemical-vapor-deposition grown bulk diamond with a ⟨001⟩ crystal orientation (electronic grade from Element Six) without any dose or annealing. A negatively charged NV center located about 30 μm below the surface was found using a confocal laser microscope. A 25-μm copper wire mechanically attached to the surface of the diamond was used to apply a microwave. An external magnetic field was applied to carefully compensate for the geomagnetic field of about 0.045 mT using a permanent magnet with monitoring of the ODMR spectrum within 0.1 MHz. The Rabi oscillation and Ramsey interference were also used to fine-tune the field. The NV center used in the experiment showed hyperfine splittings caused by $^{14}$N nuclear spin at 2.2 MHz, and by $^{13}$C nuclear spins within 0.4 MHz. All experiments were performed at 5 K to reduce the optical line width of the $A_2$ transition to as narrow as 54 MHz (Fig. 2c). The splitting between the $E_x$ and $E_y$ transitions indicated the crystal strain was 2.2 GHz in absolute value.

The experimental setup was the same as in Ref. 2, except for an additional red laser for the spin rotation (Fig. 2a). Figure 2b contains schematics of the pulse sequence used in the experiments. The arbitrary geometric spin state was prepared by the dark-state preparation method used in Ref. 2. A green laser (532nm, 100 μW) of 3 μs was first used to initialize the electron spin states to the ground $m_S = 0$ state $|0\rangle$, and then a microwave (2.878 GHz) excited the electron spin to the geometric spin state. Finally, a red light resonant to the $A_2$ state (0.8 μW, 130 ns) continued to excite the bright state until it reached the corresponding dark state. The prepared geometric spin was rotated by another red light almost resonant to the $A_2$ state (around 30 μW) in any arbitrary axis corresponding to the bright state of the light polarization, and by any arbitrary angle defined by the detuning. In this context, the optical geometric spin rotation is regarded as a bright-axis rotation. The rotated geometric spin state was then projected into any arbitrary state by the bright-state projection method used in Ref. 2. The red light (0.1 μW, 10 ns) resonantly excites the bright state to the $A_2$ state, which then relaxes to the ground state while emitting a photon with a different wavelength (phonon sideband emission) for detection. All the light beams were focused onto the sample using a 0.7 NA 100x objective inside the vacuum.

**Hamiltonian used for the analysis.** The fine structure of the orbital excited states in an NV center are well defined at temperatures below ~10 K, and the eigenstates are individually accessible with a resonant light. The Hamiltonian for the excited state is described as

$$H_{es} = \lambda_{es}^{\parallel} L_z S_z + D_{es}^{\parallel} L_z^2 S_z^2 + D_{es}^{\perp}(L_-^2 S_+^2 + L_+^2 S_-^2)$$
$$+ e_x(L_-^2 + L_+^2) + ie_y(L_-^2 - L_+^2), \tag{1}$$

where $L_z$ and $S_z$ are the axial component of the orbital and spin angular momentum operators, $L_\pm$ and $S_\pm$ are the raising and lowering operators defined as $L_\pm = (L_x \pm iL_y)/\sqrt{2}$ and $S_\pm = (S_x \pm iS_y)/\sqrt{2}$ on the SU(3) system in the bases of $\{|+1\rangle, |0\rangle, |-1\rangle\}$, $\lambda_{es}^{\parallel}$ is the axial spin-orbit interaction, $D_{es}^{\parallel}$ and $D_{es}^{\perp}$ are zero-field splittings arising from the axial and perpendicular spin-spin interaction, respectively, in the excited states, and $e_x$ and $e_y$ are the x and y components of the crystal strain. The following six eigenstates are energetically well separated except for the $E_1$ and $E_2$ states under a low strain regime.

$$|A_2\rangle = (|+1\rangle_L|-1\rangle_S + |-1\rangle_L|+1\rangle_S)/\sqrt{2} \tag{2}$$

$$|A_1\rangle = (|+1\rangle_L|-1\rangle_S - |-1\rangle_L|+1\rangle_S)/\sqrt{2} \tag{3}$$

$$|E_1\rangle = (|+1\rangle_L|+1\rangle_S + |-1\rangle_L|-1\rangle_S)/\sqrt{2} \tag{4}$$

$$|E_2\rangle = (|+1\rangle_L|+1\rangle_S - |-1\rangle_L|-1\rangle_S)/\sqrt{2} \tag{5}$$

$$|E_x\rangle = (|+1\rangle_L - |-1\rangle_L)|0\rangle_S/\sqrt{2} \tag{6}$$

$$|E_y\rangle = i(|+1\rangle_L + |-1\rangle_L)|0\rangle_S/\sqrt{2} \tag{7}$$

On the other hand, the Hamiltonian for the orbital ground state is described as

$$H_{gs} = D_{gs}S_z^2, \tag{8}$$

where $D_{gs}$ is the zero-field splitting arising from the spin-spin interaction in the ground state. The optical excitation induces the orbital transition depending on the light polarization. The driving Hamiltonian is described as

$$H_{drive} = \frac{\Omega}{\sqrt{2}}\left(\cos\left(\frac{\theta}{2}\right)|+1\rangle_L\langle 0| + e^{i\phi}\sin\left(\frac{\theta}{2}\right)|-1\rangle_L\langle 0|\right) + \text{H.c.}, \tag{9}$$

where $\theta$ ($\phi$) denotes a polar (azimuth) angle in the Poincare sphere, which represents a light polarization state, and H.c. indicates the Hermit conjugate. The driving Hamiltonian for the resonant optical transition between $|0\rangle_L|\pm 1\rangle_S$ (hereafter indicated as $|\pm 1\rangle$) and $|A_2\rangle$ is written as follows to conserve the spin angular momentum.

$$H_\Lambda = \frac{\Omega}{2}\left(\cos\left(\frac{\theta}{2}\right)|A_2\rangle\langle -1| + e^{i\phi}\sin\left(\frac{\theta}{2}\right)|A_2\rangle\langle +1|\right) + \text{H.c.}, \tag{10}$$

where $\Omega$ is the Rabi frequency. The Hamiltonian describes the dynamics in the degenerate three-level $\Lambda$ system in the computational bases.

**Bright-state driving.** The $|\pm 1\rangle$ basis states can be transformed into the bright state $|B\rangle$ and dark state $|D\rangle$, where the bright state is coupled with the excited state to create a new eigenstate, while the dark state is kept in its eigenstate. With this transformation, the driving Hamiltonian in the degenerate three-level Λ system is transformed as follows:

$$H_\Lambda = \frac{\Omega}{2}(|A_2\rangle\langle B| + |B\rangle\langle A_2|), \tag{11}$$

where

$$|B\rangle = \sin\left(\frac{\theta}{2}\right)|+1\rangle + e^{i\phi}\cos\left(\frac{\theta}{2}\right)|-1\rangle, \tag{12}$$

$$|D\rangle = \cos\left(\frac{\theta}{2}\right)|+1\rangle - e^{i\phi}\sin\left(\frac{\theta}{2}\right)|-1\rangle, \tag{13}$$

The bright state represented in the Bloch sphere shows one-to-one correspondence with the light polarization $|\psi\rangle_p$ represented in the Poincare sphere as

$$|\psi\rangle_p = \cos\left(\frac{\theta}{2}\right)|+1\rangle_p + e^{i\phi}\sin\left(\frac{\theta}{2}\right)|-1\rangle_p, \tag{14}$$

where $|\pm 1\rangle_p$ indicates right and left circular polarizations.

In general, the Λ system Hamiltonian under driving light needs to add the detuning frequency $\Delta$ as follows:

$$\begin{aligned}H_\Lambda &= \frac{\Omega}{2}(|A_2\rangle\langle B| + |B\rangle\langle A_2|) - \Delta|A_2\rangle\langle A_2| \\ &= \frac{\Omega_{\text{eff}}}{2}\boldsymbol{n}\cdot\boldsymbol{\sigma}^{\{A_2,B\}} - \frac{\Delta}{2}\sigma_0^{\{A_2,B\}},\end{aligned} \tag{15}$$

where $\Omega_{\text{eff}} = \sqrt{\Omega^2 + \Delta^2}$ is the effective Rabi frequency, $\boldsymbol{n} = \left(\frac{\Omega}{\Omega_{\text{eff}}}, 0, \frac{-\Delta}{\Omega_{\text{eff}}}\right)$ is a unit vector indicating the rotation vector, and $\boldsymbol{\sigma}^{\{A_2,B\}} = \left(\sigma_x^{\{A_2,B\}}, \sigma_y^{\{A_2,B\}}, \sigma_z^{\{A_2,B\}}\right)$, $\sigma_{0,x,y,z}^{\{A_2,B\}}$ are the Pauli operators and identity operator based on $|A_2\rangle$ and $|B\rangle$.

The time evolution operator is now written as

$$U_\Lambda(t) = \exp\left(-i\frac{\Omega_{\text{eff}}t}{2}\boldsymbol{n}\cdot\boldsymbol{\sigma}^{\{A_2,B\}}\right)\exp\left(i\frac{\Delta t}{2}\right) + |D\rangle\langle D|. \qquad (16)$$

The evolution operator for a round trip $t_{2\pi} = \frac{2\pi}{\Omega_{\text{eff}}}$ in the $|A_2\rangle - |B\rangle$ Bloch sphere becomes

$$U_\Lambda(t_{2\pi}) = \exp\left\{-i\pi\left(1 - \frac{\Delta}{\Omega_{\text{eff}}}\right)\right\}\sigma_0^{\{A_2,B\}} + |D\rangle\langle D|. \qquad (17)$$

The operation is interpreted as the rotation around the bright state $|B\rangle$ by an angle $\pi\left(1 - \frac{\Delta}{\Omega_{\text{eff}}}\right)$ in the Bloch sphere spanned with the $|\pm 1\rangle$ basis states. An arbitrary angle rotation around an arbitrary axis thus becomes possible by choosing the proper light polarization, a detuning $\Delta$, and the corresponding pulse length.

**Hamiltonian compensation on the light polarization.** Since NV centers in diamond oriented along <100> are off-aligned from the optical axis by $\arccos(\frac{1}{\sqrt{3}})$, the electric-field amplitude of the light polarized along the inclined NV axis is decreased by a factor of $1/\sqrt{3}$ as the electric field is projected onto the plane normal to the NV axis, resulting in a reduction of the Rabi frequency. As a result, the polarization observed by the NV electron is different from the incident polarization due to the off-alignment. In addition, the crystal strain mixes the $|A_2\rangle$ state with the $|E_1\rangle$ and $|E_2\rangle$ states to rearrange the straightforward correspondence between the bright state, defined as Eq. (12), and the light polarization, defined as Eq. (14).

In this experiment, we estimated the strain parameters as $E_x = -1.2$ GHz, $E_y = -1.8$ GHz from the fitting to the quantum state tomography of the spin states rotated around the x-, y- and z-axes. As in Eq. (15), the ideal Hamiltonian is then created to calibrate both the off-alignment and strain effects by finding the adopted light polarization. The bright and dark states are obtained by projecting the eigenstates of the total Hamiltonian as the sum of Eqs. (1), (8) and (9) onto the qubit space $|\pm1\rangle$. The light polarization adapting to the Hamiltonian can be found to realize the expected rotation. The light polarization parameters with and without the compensation for the $\pm$X-, $\pm$Y- and $\pm$Z-rotations are summarized in Supplementary Table 1.

**Master equation.** We analyzed the experimental data shown in Figs. 2d and 4a based on the Lindblad master equation to simulate the relaxation process. The relaxation is caused by two processes: energy relaxation with a decay time of $T_1$ = 12 ns of the excited state $|A_2\rangle$ to the $|\pm1\rangle$ state, and phase relaxation with a dephasing time of $T_2^*$ between the $|\pm1\rangle$ states. The $T_2^*$ was estimated to be 4.6 ns from the fitting to the Rabi oscillation shown in Fig. 2d. We neglected the relatively long energy relaxation from $|A_2\rangle$ to the $|0\rangle$ for simplicity. The phase relaxation between the $|\pm1\rangle$ states was also negligible, since the dephasing time was on the order of a few μs.

**Quantum process tomography.** The $\chi$ matrices shown in Fig. 3c to represent the rotation gate operations in Fig. 3b were reconstructed via quantum process tomography[30], which compares the final state after the gate operation with the initial state prepared in the $|+\rangle, |+i\rangle, |+1\rangle, |-1\rangle$. Since the obtained raw $\chi$ matrices were likely to be unphysical, the most likely $\chi$ matrices were deduced by assuming the trace conservation and the normal matrix based on the maximum likelihood estimation method[32]. The optimization was performed by using the differential_evolution function, which enables global optimization, in the scipy.optimize package of the python program.

**Data availability.** The data that support the plots within this paper and other findings of this study are available from the corresponding author upon reasonable request.

## Acknowledgments

We thank Yuichiro Matsuzaki, Burkhard Scharfenberger, Kae Nemoto, William Munro, Norikazu Mizuochi, Nobuyuki Yokoshi, Fedor Jelezko, and Joerg Wrachtrup for their discussions and experimental help. This work was supported by National Institute of Information and Communications Technology (NICT) Quantum Repeater Project, and by Japan Society for the Promotion of Science (JSPS) Grant-in-Aid for Scientific Research (24244044, 16H06326, 16H01052) and Ministry of Education, Culture, Sports, Science and Technology (MEXT) as "Exploratory Challenge on Post-K computer" (Frontiers of Basic Science: Challenging the Limits).

## Author Contributions

N.N. carried out the experiment. Y.S, R.K. and H.Kano supported the experiment. Y.S. and H.Kosaka analysed the data. Y.S. and H.Kosaka wrote the manuscript. H.Kosaka supervised the project. All authors discussed the results and commented on the manuscript.

## Additional Information

Correspondence and requests for materials should be addressed to H.Kosaka.

## Competing Financial Interests statement

The authors declare no competing financial interests.26

# Figures

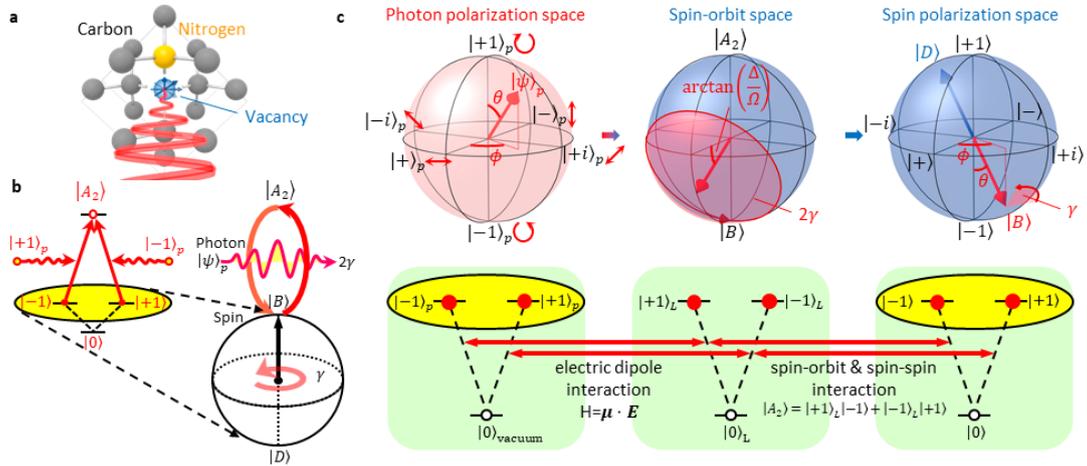

**Figure 1 | Optical geometric spin rotation.** (a) Molecular structure of a nitrogen vacancy in diamond. (b) Degenerate three-level Λ system and conceptual explanation for the geometric spin rotation. (c) Correspondence between the polarization vector for the rotation light represented in the Poincare sphere (left) and the rotation vector for the geometric spin represented in the Bloch sphere (right). The rotation angle $\gamma$ is determined by the solid angle $2\gamma$ of the cyclic evolution in the spin-orbit space based on the relevant excited state $|A_2\rangle$ and the bright state $|B\rangle$ for the rotation light (middle).

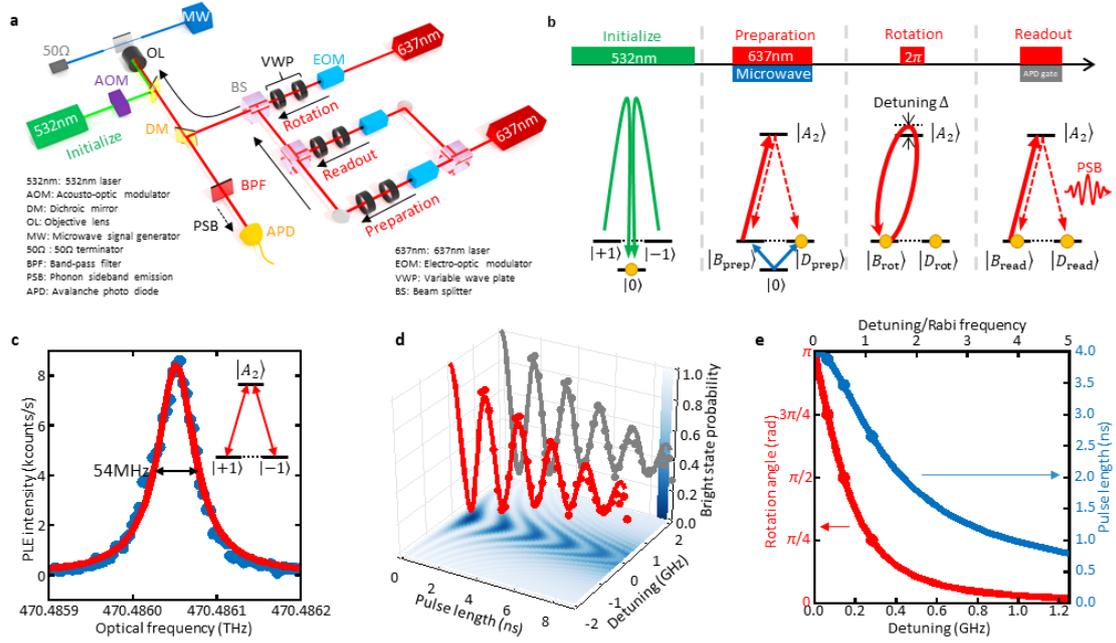

**Figure 2 | Experimental procedure and characterization of the NV center.** (a) Experimental setup. (b) Experimental pulse sequence and related transition in the three-level Λ system. The degenerate $|\pm 1\rangle$ states are reconfigured into bright $|B\rangle$ and dark $|D\rangle$ states defined by the polarizations for the preparation, rotation, and readout lights. (c) The photoluminescence excitation (PLE) spectrum with respect to the optical transition from the $|\pm 1\rangle$ states to the $A_2$ state, which has a lifetime of 12 ns. (d) Optically-driven Rabi oscillation for the vertically polarized light $|-\rangle_p = (|+1\rangle_p - |-1\rangle_p)/\sqrt{2}$ in the spin-orbit space. The vertical axis indicates the bright-state population. (e) Geometric rotation angle (red) and $2\pi$-pulse length in the spin-orbit space (blue) as a function of the detuning frequency from the $A_2$ resonance, where the Rabi frequency on resonance is 250MHz.

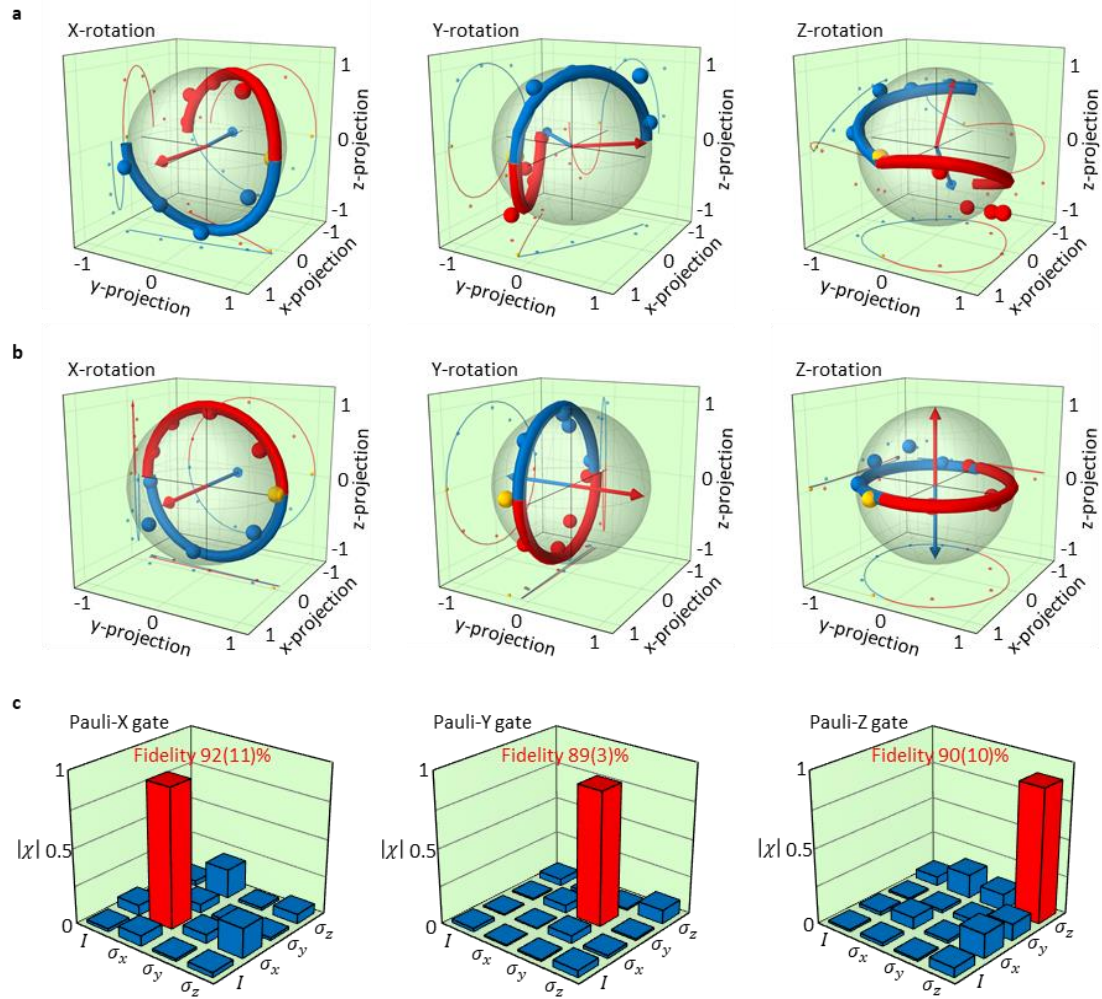

**Figure 3 | Optical geometric spin rotation.** (a) The geometric spin states rotated about the X-, Y- and Z-axes, beginning at the $|+i\rangle$, $|+\rangle$ and $|+\rangle$ states, respectively (yellow dots), are plotted in the Bloch sphere. The red (blue) dots show the states rotating around the positive (negative) bright state vector (indicated with arrows). Solid lines show the fitting curves, assuming the X and Y components of the off-alignment of the NV center and the crystal strain based on the Hamiltonian analysis. (b) The same plots after polarization compensation for the state preparation, rotation and readout. The

arrows indicate the bright states that represent the rotation axes. (c) χ matrix elements

reconstructed by the quantum process tomography to represent the optical holonomic

Pauli-X, Y and Z gate operations after the polarization compensation. Numbers in parentheses indicate the standard deviations of the experimentally derived matrix elements from those estimated by the maximum likelihood method.

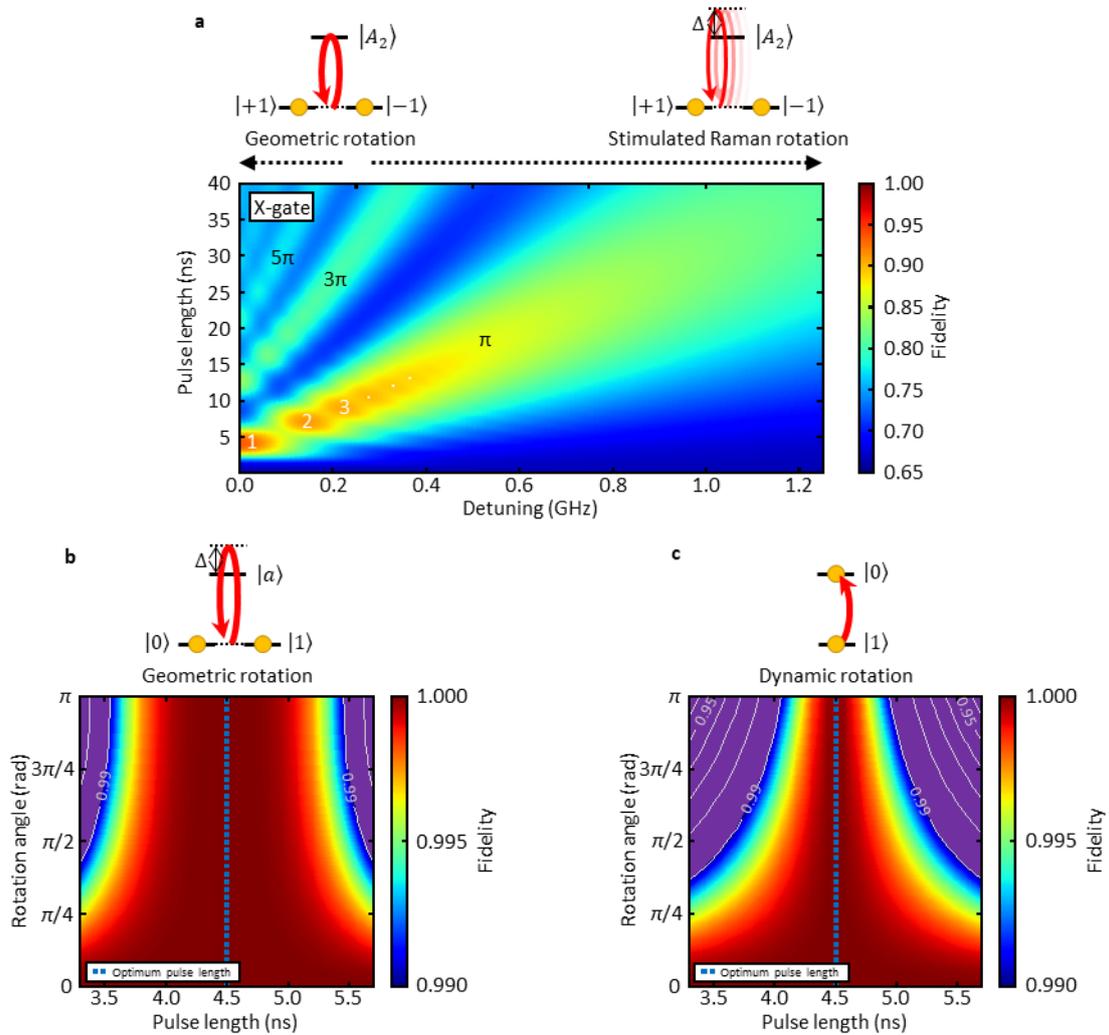

**Figure 4 | Simulated fidelity tolerance of the holonomic quantum gates.** (a) The

dependence of fidelity on detuning for X-gates considering relaxation using the Master equation based on parameters obtained from the experiments shown in Fig. 2d. White numbers indicate the number of turns used to perform the X-gates. Black numbers indicate spin rotation angles. (b, c) Fidelity tolerance of (b) geometric rotations for the three-level $\Lambda$ system consisting of degenerate basis states $\{|0\rangle, |1\rangle\}$ and an ancillary state $|a\rangle$; and (c) dynamic rotations for the two-level system consisting of gapped basis states $\{|0\rangle, |1\rangle\}$ against pulse length without considering relaxation.